\documentclass[prd,aps, preprintnumbers, showpacs,twocolumn, nofootinbib,superscriptaddress,notitlepage]{revtex4-1}
\usepackage[normalem]{ulem}
\usepackage{epsfig}
\usepackage{amsfonts}
\usepackage{amsmath}
\usepackage{slashed}
\usepackage{graphicx}
\usepackage{color}
\usepackage{mathtools}
\usepackage{subfigure}
\usepackage{graphicx}

\usepackage{stmaryrd}
\usepackage{amssymb,psfrag}
\usepackage[normalem]{ulem}
\usepackage{caption}

\newcommand{\beq}{\begin{eqnarray}}
\newcommand{\eeq}{\end{eqnarray}}

\newcommand{\nn}{\nonumber}

\begin{document}

\title{Nucleons as modified Ising models}

\author{Shu-Man Hu}
\email{hsm1232022@163.com}
\affiliation{School of Physics and Microelectronics, Zhengzhou University, Zhengzhou, Henan 450001, China}

\author{Yin-Sen Luan}
\email{Corresponding author. luanyinsen@gmail.com}
\affiliation{PLA Strategic Support Force Information Engineering University, Zhengzhou 450001, China}
\affiliation{School of Mechanical Engineering, Shanghai Jiao Tong University, Shanghai 200240, China}

\author{Ji Xu}
\email{Corresponding author. xuji\_phy@zzu.edu.cn}
\affiliation{School of Physics and Microelectronics, Zhengzhou University, Zhengzhou, Henan 450001, China}

\begin{abstract}
In this paper, we propose a map which connects nucleons bound in nuclei and Ising spins in Ising model. This proposal is based on the fact that the description of states of nucleons and Ising spins could share the same type of observables. We present a nuclear model as a correspondence to an explicit modified Ising model and qualitatively confirm the correctness of this map by simulation on a two-dimensional square lattice. This map would help us understand the profound connections between different physical systems.
\end{abstract}

\maketitle

\section{Introduction}
Inspired by the renowned fermion-boson duality and recently proposed fermion-bit correspondence\,\cite{Girardeau:1960,Wetterich:2016yaw}, we propose a map between nucleons which bound in nuclei and Ising spins in Ising model, it is labeled as ``SRC-bit map'', here SRC is abbreviation of short-range correlation. Brief introduction of these sorts of maps would be presented subsequently.

The fermion-boson duality shows the equivalence of fermionic and bosonic particle systems, it stems from the following question --- can bosons and fermions transform into each other? It is possible in supersymmetry, a hot candidate for solving the hierarchy problem, which is still waiting to be examined by high energy experiment. However, in low dimensional non-relativistic systems, the equivalence of bosonic and fermionic systems has been reported long time ago\,\cite{Girardeau:1960,Girardeau:1965,Mattis1965,Coleman:1974bu,Tomonaga:1950zz,Luttinger:1963zz,schmidt:1998,crescimanno:2001,cheon:pla,Cheon:1998iy}. Massless boson and fermion theories in $1+1$ dimensional Minkowski and curved space-time are proved to be equivalent\,\cite{freundlich:1972,davies:1978}. Because the spin-statistics relation is based on relativity, non-relativistic system can escape from the spin-statistics relation, thus boson can be spinning and fermion be spinless. The relation between  boson and spinless fermion may shed light on the general properties of boson-fermion duality.

Fermion-bit map is proposed based on the equivalence between the two formulations for describing fermions and Ising spins. They could have the identical expectation values of products of observables. In the case of fermions, the observables are occupation numbers $n(x)$ that take values 0 or 1. For $n(x)=1$ a fermion is present at $x$, while for $n(x)=0$ no fermion located in $x$. In the case of Ising spins, $s(x)$ can take values $\pm 1$ which can be understood as magnetic dipole moments of atomic ``spins'' in ferromagnetism. A relation between occupation numbers and Ising spins can be readily established, $n(x)=(s(x)+1)/2$. Based on these simple observations, a map between fermions and Ising models has been proposed\,\cite{Wetterich:2009tq,Wetterich:2010eh}. Since Ising spins can be associated to bits of information, this map is under the name of ``fermion-bit map''. It is worth noting that if Ising spins are considered as ``discrete bosons'', this fermion-bit map establishes a general equivalence of fermions and discrete bosons.

Inspired by these relations, we propose a new one between nucleons and Ising spins. One may think of the stable nucleus as a tight ball of neutrons and protons (collectively called nucleons), held together by the strong nuclear force. This basic picture has been worked very well. Afterwards, deep inelastic scattering (DIS) led to the discovery that the nucleon is made of quarks. However, due to the small nuclear binding energy and the idea of quark-gluon confinement, it was thought that quarks had no explicit role in the nucleus, hence nuclei could still be described in terms of nucleons and mesons. In 1982, this understanding was changed by measurements performed by the European Muon Collaboration (EMC)\,\cite{Aubert:1983xm}. The initial expectation was that physics at GeV scale would be insensitive to the nuclear binding effects which are typically on the order of several MeV scale. However, the collaboration discovered the per-nucleon deep inelastic structure function in iron is smaller than that of deuterium in the region $0.3<x_B<0.7$, here $x_B$ is the Bjorken variable. This phenomenon is known as EMC effect and has been observed for a wide range of nuclei\,\cite{Arneodo:1988aa,Arneodo:1989sy,Allasia:1990nt,Gomez:1993ri,Seely:2009gt}. Although the understanding of how the quark-gluon structure of a nucleon is modified by the surrounding nucleons has been brought to a whole new level, there is still no consensus as to the underlying dynamics that drives this effect. Currently, one of the leading approaches for describing the EMC effect is: nucleons bound in nuclei are unmodified, same as ``free'' nucleons for most of the time, but are modified substantially when they fluctuate into SRC pairs. The connection between SRC and EMC effects has been extensively investigated in nuclear structure function measurements\,\cite{Egiyan:2005hs,Hen:2012fm,Hen:2014nza,Duer:2018sby,Weinstein:2010rt,Chen:2016bde,Lynn:2019vwp,Xu:2019wso,Hatta:2019ocp,Huang:2021cac,Hen:2016kwk}.
SRC pairs are conventionally defined in momentum space as a pair of nucleons with high relative momentum and low center-of-mass (c.m.) momentum, where high and low are relative to the Fermi momentum of medium and heavy nuclei. In this paper, we will emphasize the similarities between the descriptions of SRCs and Ising spins. Based on these similarities, a new kind of map has been proposed in this paper, it is labeled as ``SRC-bit map''.

The present manuscript is arranged as follows. In Sec.\,\ref{duality_s_i}, we will discuss the map between SRCs in nucleons and Ising spins. One longstanding theoretical model which is used to depict the nucleons is presented as a reference to an explicit Ising model. Sec.\,\ref{Simulation} devotes into simulations of nucleon states in terms of Ising model, our preliminary results support the proposed map between nucleons and Ising spins. Finally, we summarize our work and comment on future developments in Sec.\,\ref{conclusions}.

\section{Map between SRCs and Ising spins}
\label{duality_s_i}
\subsection{Notations and definitions}
\label{duality_s_i_1}

\begin{figure}[htbp]
\centering
\includegraphics[width=0.98\columnwidth]{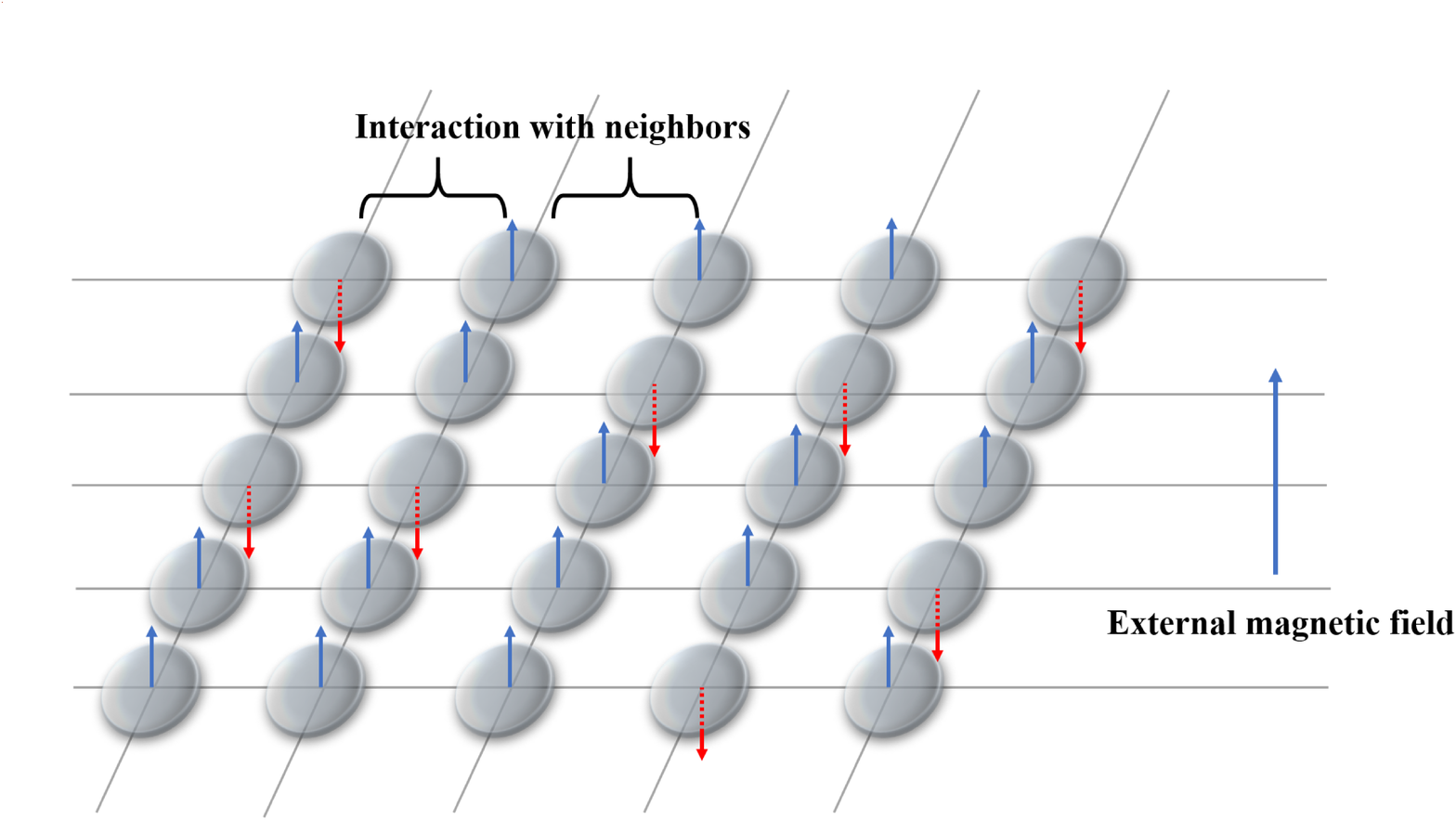}
\centering
\includegraphics[width=0.98\columnwidth]{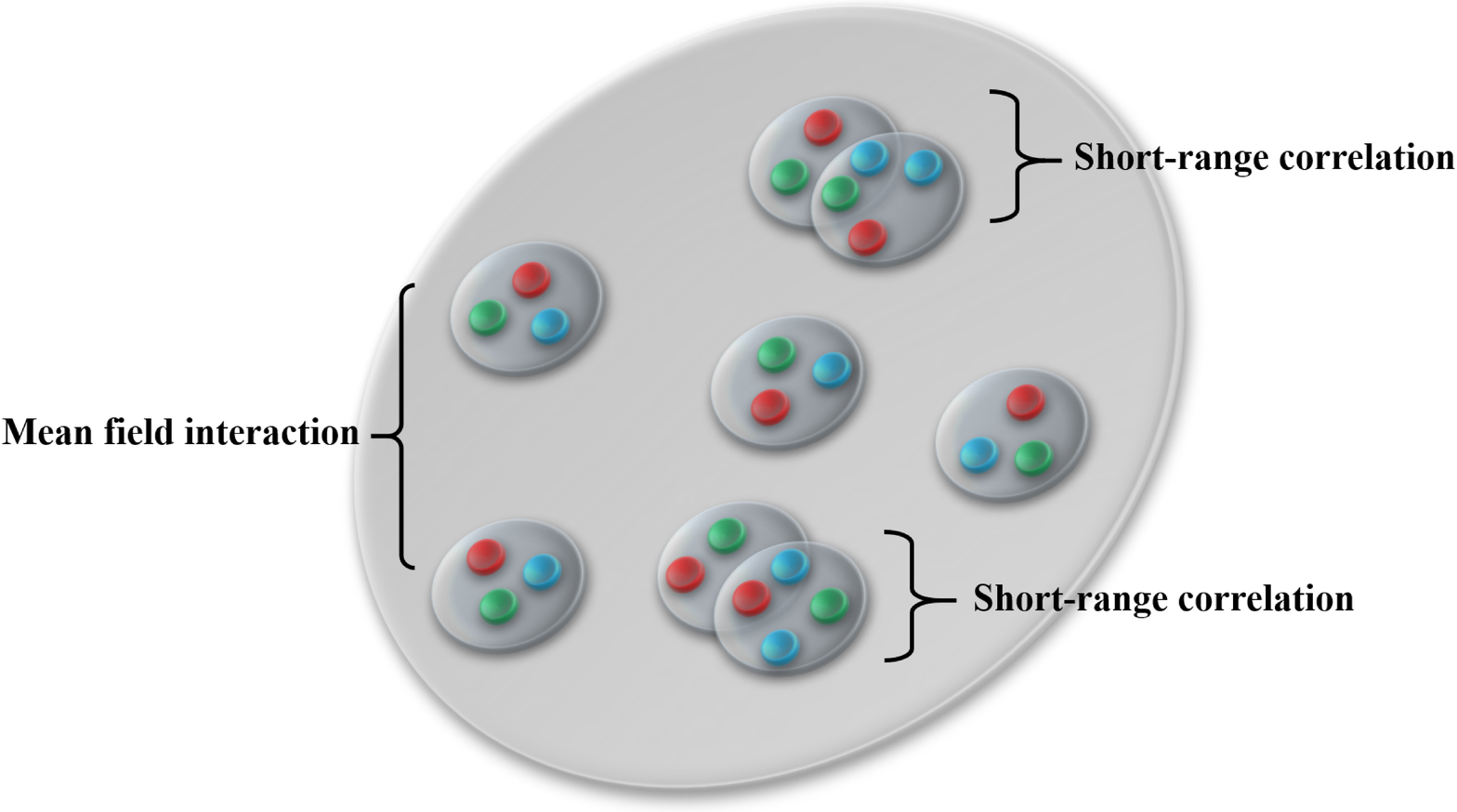}
\centering
\caption{Schematic diagrams for Ising model in ferromagnetism (upper) and nucleus structure (lower), both of the types of interactions in these diagrams can be divided into two parts. For Ising model, they are adjacent interaction and universal external magnetic field interaction. For nucleus structure, they are SRCs and interaction with mean field.}
\label{picsIS}
\end{figure}

The description of SRCs in nucleons and Ising spins in Ising model could share the same type of observables, as in line with implementations of fermions within Ising model. Ising spin $s(x)$ can take values $\pm 1$, each represents one of two spin states (spin up or down). Similarly, we can use $s_{src}(x)$ to represent the state of a nucleon. In nuclei, nucleons behave approximately as independent particles in a mean field, but occasionally ($20\%-25\%$ in medium or heavy nuclei) two nucleons get close enough to each other so that temporarily their singular short-range interaction cannot be well described by a mean-field approximation. These are the two-nucleon short-range correlations (2N-SRC). For $s_{src}(x)=1$ a nucleon which belongs to SRC pair is presented, while for $s_{src}(x)=0$ the corresponding nucleon can be regarded as independent particle. The relation between these two ``spins'' is
\begin{eqnarray}\label{relation}
  s(x)=2s_{src}(x)-1 \,.
\end{eqnarray}
In addition to this simple relation, there are other common features between these two systems. In Ising model, the spins are arranged in a lattice, allowing each spin to interact with its neighbors. Consider a set of lattice sites $\Lambda$, for each lattice site $x\in \Lambda$ there is a discrete variable $s(x)$ such that $s(x)\in\{+1,-1\}$, representing the site's spin. For any two adjacent sites $x,y\in\Lambda$ there is an interaction $J$. Besides, every site $y\in\Lambda$ is influenced by an external magnetic field $h$, the corresponding Hamiltonian is
\begin{eqnarray}\label{Ising_H}
  H = -\frac{J}{2}\sum_{\langle x y\rangle} s_{x} s_{y}- \mu h \sum_{y} s_{y} \,,
\end{eqnarray}
where the first sum is over pairs of adjacent spins and the second term represents the universal interaction with external magnetic field, the magnetic moment is given by $\mu$. The physical quantities describing nucleons can also be divided into two parts, one for short-range correlation state and the other for long-range state. The pedagogically sketched diagrams for Ising model and structure of nucleus are presented in Fig.\,\ref{picsIS}. For instance, the nucleon spectral function $P(\mathbf{p},E)$ which is the joint probability to find a nucleon in a nucleus with momentum $\mathbf{p}$ and removal energy $E$ can be modeled as\,\cite{CiofidegliAtti:1995qe}
\begin{eqnarray}
  P(\mathbf{p}, E)= P_{1}(\mathbf{p}, E) + P_{0}(\mathbf{p}, E) \,,
\end{eqnarray}
where the subscript $1$ refers to high-lying continuum states that are caused by the short-range correlations and the subscript $0$ refers to values of $E$ corresponding to low-lying intermediate excited states.

Another example is nuclear gluon distribution, we can parameterize the nuclear gluon distribution in the EMC region as that for the structure function\,\cite{Xu:2019wso,Frankfurt:1993sp,Segarra:2019gbp}
\begin{eqnarray}
  g_{A}\left(x_B, Q^{2}\right)= 2 n_{src}^{A} \delta \tilde{g}\left(x_B, Q^{2}\right) +A g_{p}\left(x_B, Q^{2}\right) \,,
\end{eqnarray}
where $n_{src}^{A}$ represents number of SRC pairs in nucleus $A$. Here the authors have made an approximation that all nuclear modifications originate from the nucleon-nucleon SRCs in the EMC region. $\delta \tilde{g}\left(x_B, Q^{2}\right)$ represents the difference between the gluon distribution in the SRC pair and in the free nucleon.

Inspired by the term ``fermion-bit map'' presented in Ref.\,\cite{Wetterich:2016yaw}. We propose a ``SRC-bit map'' which represents the relation between the states of each nucleon in nucleus and the state of each lattice site in Ising model. An obvious benefit of this map is that it allows to describe properties of nucleons in terms of classical statistical systems for Ising spins, with many highly developed methods available. The two-nucleon short-range correlations are defined operationally in experiments as having small center-of-mass (c.m.) momentum and large relative momentum, to which there are approximately $20\%$ nucleons belong. This means in a medium or heavy nuclei, the average $\bar{s}_{src}=0\times 80\%+1\times 20\%=0.2$, which corresponds to average Ising spin $\bar{s}=-0.6$ in terms of Eq.\,(\ref{relation}).

\begin{figure}[htbp]
\centering
\includegraphics[width=0.5\columnwidth]{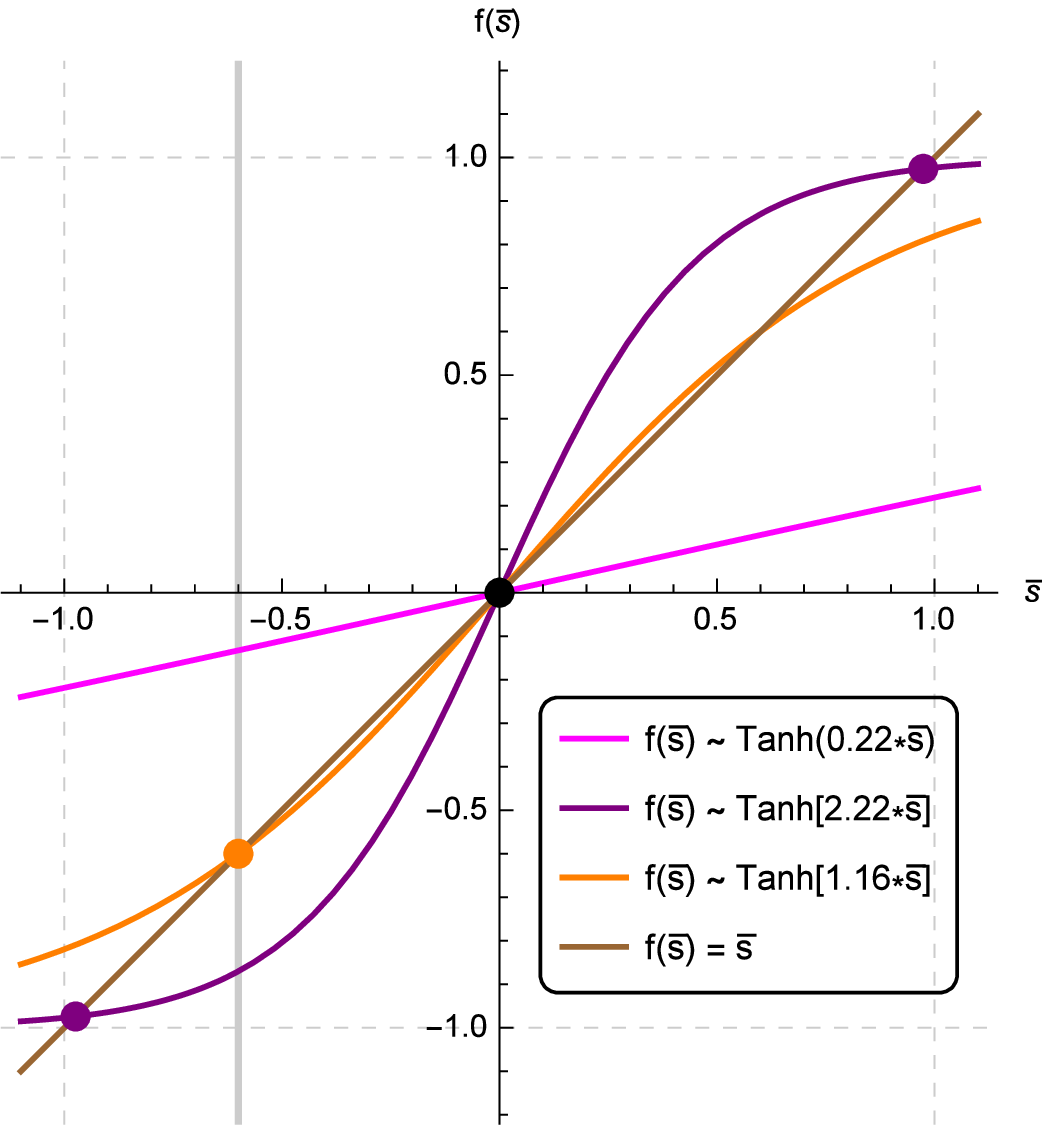}
\centering
\hspace{15mm}
\includegraphics[width=0.5\columnwidth]{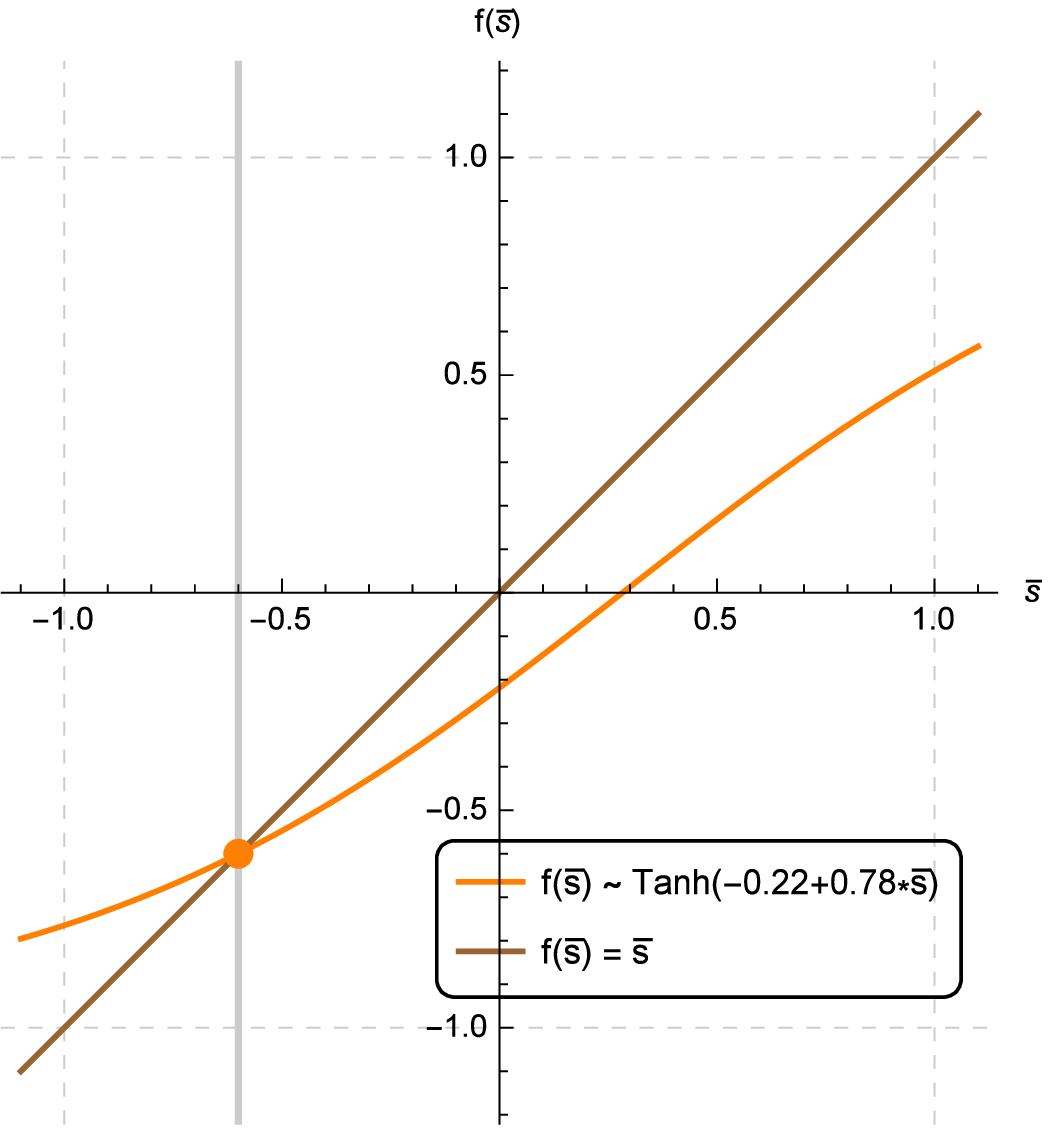}
\centering
\caption{Upper panel shows the shapes of r.h.s. of Eq.\,(\ref{nohsimp}) with different coefficients, the brown one is $f(\bar{s})=\bar{s}$. Desired solution (orange point) could be acquired with appropriate choice of coefficient $(\mathcal{Z} J)/(k_{B} T)$. Lower panel shows the typical shape of r.h.s. of Eq.\,(\ref{bars}) with desired solution (orange point).}
\label{pics12}
\end{figure}

For a Hamiltonian shown in Eq.\,(\ref{Ising_H}), the partition function is
\begin{eqnarray}
  Z=\sum_{\left\{s(x)\right\}} e^{-\beta H} \,,
\end{eqnarray}
where $\beta=(k_B T)^{-1}$ and $\left\{s(x)\right\}$ means sum over all possible configurations of spins. If the approach of mean field approximation is utilized, the Hamiltonian can be further simplified,
\begin{eqnarray}
  H_{MF}=-\sum_{x} \mu s(x)\left(h+\bar{h}\right) \,,
\end{eqnarray}
here $\bar{h}\equiv (\mathcal{Z} J/\mu)\bar{s}$, $\mathcal{Z}$ is known as coordination number and $\bar s$ is average Ising spin. Therefore, the partition function reads
\begin{eqnarray}
Z_{MF} &=& \prod_{x=1}^{N}\left(\sum_{s(x)=\pm 1} e^{-\beta \mu(h+\bar{h}) s(x)}\right) \nn\\
&=& \left[2 \cosh \left(\frac{\mu h}{k_{B} T}+\frac{\mathcal{Z} J }{k_{B} T}\bar{s}\right)\right]^{N} \,.
\end{eqnarray}
The magnetization can be written as
\begin{eqnarray}\label{M12}
 \left\{\begin{array}{l}
  M=N \mu \bar{s} \,,\\
  M=-(\partial F/\partial h)=N\mu \tanh \left( \frac{\mu h}{k_{B} T}+\frac{\mathcal{Z} J }{k_{B} T}\bar{s} \right) \,,
 \end{array}\right.
\end{eqnarray}
here $F$ is the Helmholtz free energy. $\bar s$ can be deduced from Eq.\,(\ref{M12})
\begin{eqnarray}\label{bars}
  \bar{s}=\tanh \left(\frac{\mu h}{k_{B} T}+\frac{\mathcal{Z} J}{k_{B} T} \bar{s}\right) \,.
\end{eqnarray}
Recall that $\bar{s}=-0.6$ is the counterpart of $\bar{s}_{src}=0.2$. It is interesting to find out under which condition can we get $\bar{s}=-0.6$. First, in the case of Ising model with no external magnetic field, Eq.\,(\ref{bars}) simplifies
\begin{eqnarray}\label{nohsimp}
  \bar{s}=\tanh \left(\frac{\mathcal{Z} J}{k_{B} T} \bar{s}\right) \,.
\end{eqnarray}
It is the coefficient $(\mathcal{Z} J)/(k_{B} T)$ determines the solution. The upper panel of Fig.\,\ref{pics12} shows shapes of r.h.s. of Eq.\,(\ref{nohsimp}) with different coefficients, one can see that the wanted solution $\bar{s}\!\! = \!\!-0.6$ can be achieved when $(\mathcal{Z} J)/(k_{B} T)\!\! \approx \!\!1.16$. Second, in the case of Ising model with external magnetic field, both $(\mu h)/(k_B T)$ and $(\mathcal{Z} J)/(k_{B} T)$ will contribute to the result, the lower panel of Fig.\,\ref{pics12} presents the typical shape of r.h.s. of Eq.\,(\ref{bars}) with solution $\bar{s}=-0.6$.

We would like to give a few remarks here:
\begin{enumerate}
  \item All these discussions are independent of the particular dynamics of the systems. The Sect.\,\ref{duality_s_i} in our paper therefore constitutes a very general map from a discrete classical statistical ensemble to structure of nucleus which is formed by strong interaction. In this respect it is in line with earlier implementations of fermions within bosonic systems under gauge transformation.
  \item According to the simple relation in Eq.\,(\ref{relation}), there must be phase transition in Ising model in order to map the SRC phenomenon in nucleus (since $\bar{s}=0$ leads to $\bar{s}_{src}=0.5$, this is not the correct number measured in experiments). In one dimension, the solution of Ising model admits no phase transition, thus one should go to higher dimension to delve into this subject rigorously.
  \item There are some arguments that single-particle correlations such as momentum distributions and single-particle spectral densities are not forced to be identical between bosons and fermions in Ref.\,\cite{Sekino:2020urg}. Similarly, whether SRC-bit map would help people understand parton distribution functions such as gluon distribution needs further investigation.
  \item SRC of more than two nucleons such as 3N-SRC also exists in nuclei although its probability is expected to be significantly smaller than the 2N-SRC. Thus there would be three states of a nucleon --- does not belong to SRC, belongs to 2N-SRC or 3N-SRC. The three-states Potts model is a natural extension of the Ising model where the spin on a lattice takes one of three possible values \,\cite{Wu:1982ra}. It would be intriguing to generalize the SRC-bit map into these three states situations.
\end{enumerate}

The important aspect of Ising model is that a variety of problems can be investigated by the similar kind of modeling. Undoubtedly, with many highly developed methods available for Ising model, the SRC-bit map provides a new insight into the study of EMC effect in nuclear physics.

\subsection{Potential theoretical model which can be simulated with Ising model}
\label{duality_s_i_2}
In the rest of this section, we will discuss a theoretical model as a potential candidate for possessing an explicit Ising model and thus could be investigated in terms of many highly developed statistical methods. In this theory, the nucleon can be regarded as a superposition of two different configurations where one is ``blob like'' configuration (BLC) with the normal nucleon size and the other is ``pointlike'' configuration (PLC) \,\cite{Frankfurt:1985cv}. The BLC can be thought of as an object that is similar to a nucleon. The PLC represents a three-quark system of small size which dominates the high-$x_B$ behavior of parton distribution function.

The idea that different constituents of the nucleon have different sizes is directly related to EMC effect\,\cite{Frank:1995pv}. The Hamiltonian is given by the matrix
\begin{eqnarray}
  H_{0}=\left[\begin{array}{cc}
E_{B} & V \\
V & E_{P}
\end{array}\right] \,,
\end{eqnarray}
where $E_P$ and $E_B$ are energies of PLC and BLC respectively, $V$ is the hard-interaction potential that connects the two components. We choose $E_P \! \gg \! E_B$ and $|V| \! \ll \! E_{P}-E_{B}$, so that the nucleon is mainly BLC. When placed in a nucelus, the BLC component of a nucleon feels an attractive nuclear potential Hamiltonian $H_1$,
\begin{eqnarray}\label{introU}
  H_{1}=\left[\begin{array}{ll}
U & 0 \\
0 & 0
\end{array}\right] \,.
\end{eqnarray}
Therefore the complete Hamiltonian $H \! = \!H_0+H_1$ is presented as
\begin{eqnarray}
  H=\left[\begin{array}{cc}
E_{B}-|U| & V \\
V & E_{P}
\end{array}\right] \,.
\end{eqnarray}
It is worth noting the inclusion of $U$ increase the energy difference between the BLC and the PLC components, which decreases the PLC probability. The eigenstates of $H$ are labeled as $| N \rangle_M$ and $| N^* \rangle_M$, here the subscript $M$ means medium-modified, they are approximately
\begin{eqnarray}\label{decom_NM1}
|N\rangle_{M} &=& |B\rangle+\epsilon_{M}|P\rangle \,, \nn\\
\left|N^{*}\right\rangle_{M} &=& -\epsilon_{M}|B\rangle+|P\rangle \,,
\end{eqnarray}
where $\epsilon_{M} \!=\! V/(E_{B}-|U|-E_{P})$, $|B\rangle$ stands for BLC state and $|P\rangle$ for PLC state. One can also write down the eigenstates of $H_0$,
\begin{eqnarray}
|N\rangle &=& |B\rangle+\epsilon|P\rangle \,,\nn\\
\left|N^{*}\right\rangle &=& -\epsilon|B\rangle+|P\rangle \,,
\end{eqnarray}
with $\epsilon \!=\! V/(E_B-E_P)$. Therefore the medium-modified nucleon $|N\rangle_{M}$ could be expressed in terms of the unmodified eigenstates $|N\rangle$ and $|N^*\rangle$ as
\begin{eqnarray}
  |N\rangle_{M} \approx |N\rangle+\left(\epsilon_{M}-\epsilon\right)\left|N^{*}\right\rangle \,.
\end{eqnarray}
It is the second term whose functionality resembles the SRC pair described above which dominates the high-$x_B$ behavior of structure function, ie., the EMC effect measured in DIS experiments. By adjusting the amount of excited state $\left|N^{*}\right\rangle$ contained in nucleon $|N\rangle_{M}$, the deviation of the EMC ratio from unity could be predicted\,\cite{Frank:1995pv,CiofidegliAtti:2007ork}. In this theory, the degree of deviation is controlled by $U$, $V$ and $E_P-E_B$. We use $s_{src}=1$ to represent the excited state $\left|N^{*}\right\rangle$ and $s_{src}=0$ for the ground state $\left|N\right\rangle$. According to the simple relation in Eq.\,(\ref{relation}), their correspondences are lattice sites with spin $s=1$ and $s=-1$ respectively. For an Ising model in certain dimension, the variables which determine the final magnetization state are temperature $T$, coupling $J$ and external magnetic field $h$.

One can envision the following situation, before bound in a nucleus, the nucleons can be regarded as a collection of "free particles" whose components are $|N\rangle \!=\! |B\rangle+\epsilon|P\rangle$. It is the potential $V$ that connects the two components, the amount of PLC decreases with the increase of $V^{-1}$ in which case most would be BLC. This is very similar to Ising model in ferromagnetism without external magnetic field, the states of lattices tends to be the same with the increase of coupling $J$. The energy difference $\Delta E=E_P-E_B$ is also an important factor, the number of BLC and PLC would be approximately equal when $\Delta E^{-1} \to \infty$, analogous to the case of no phase transition when $T\to\infty$ which is provided as an acceptance criteria for different spin states in Ising model. For $\Delta E^{-1} \to 0$, the huge energy difference indicates there is few PLC in nucleon, corresponding to no SRC pair. Similarly, when $T\to 0$, nearly all of the spin states are the same in Ising model.

Now suppose the nucleons are bound in a nucleus, they would feel an attractive nuclear potential $U$, which further decreases the PLC probability according to Eq.\,(\ref{decom_NM1}). Similar situation occurs on paramagnetics when we add an downward external magnetic field $h$, this would convert more spin states of lattice sites to $-1$. From this point of view the general properties of variables used to describe the models for nucleons and Ising spins are the same.

\section{Monte Carlo simulations of nucleon states}
\label{Simulation}
\subsection{Notations and definitions}
We will explore the issue of simulation on nucleon states in nucleus in terms of mature methods available for two-dimensional square lattice Ising model in more detail. In general, a Monte Carlo simulation processes a subset of configurations in the configuration space of a given
system, according to a predefined probability distribution. Here we would set the states of all lattice sites to $-1$ as our predefined distribution. Eq.\,(\ref{relation}) allows us to relate the $s=-1$ to $s_{src}=0$ of nucleon state, which corresponds to the situation where all the nuclei do not belong to SRC.

The Metropolis algorithm will be utilized to perform importance sampling of the configuration space\,\cite{CasquilhoCamp}. In this method, a Markov chain of configurations is generated in which each configuration $C_{\ell+1}$ is obtained from the previous one $C_{\ell}$ with a suitably chosen transition probability $\omega_{(\ell \to \ell+1)}$ which is determined by Metropolis function
\begin{eqnarray}\label{MProb}
  \omega_{(\ell \to \ell+1)}=\min \left[1, \exp \left(-\frac{\Delta E}{k_{B} T}\right)\right] \,,
\end{eqnarray}
ie.,
\begin{eqnarray}
\left\{\begin{aligned}
  & \omega_{(\ell \to \ell+1)} = \exp \left(-\frac{\Delta E}{k_{B} T}\right)\,, & \quad \text { if } \Delta E>0 \,,\\
  & \omega_{(\ell \to \ell+1)} = 1\,, & \quad \text { if } \Delta E<0  \,.\nn
\end{aligned}\right.\\\nn
\end{eqnarray}
here $\Delta E = E(C_{\ell+1}) - E(C_{\ell})$. The process of $C_{\ell} \to C_{\ell+1}$ constitutes one Monte Carlo step (MCS), which may be taken as our unit of computational ``time''. We will take the time to $1\times 10^6$ in our simulation.

\subsection{Modified Hamiltonian of Ising model}
The Hamiltonian of Ising model has already been shown in Eq.\,(\ref{Ising_H}). We will modify this Hamiltonian to make it more suitable for describing nucleons. The states of nucleons are simulated on a $400\times400$ lattice and these $1.6\times 10^5$ lattice sites are grouped into $8\times 10^4$ pairs taking into account the SRC always appears in pair. The spin of any pair of sites is either $+1$ or $-1$, the value of coupling $J$ depends on the magnetization state of the system.

We divide the first term in Eq.\,(\ref{Ising_H}) which describes interaction between adjacent spins into two parts, one is responsible for the adjacent interaction between a pair of $+1$ states, the other accounts for the interaction between a pair of $-1$ states. Therefore, the modified Hamiltonian of Ising model is
\begin{eqnarray}\label{M_Ising_H}
  H &=& -2C \, J_{\textrm{up}}\sum_{\langle i \rangle} s_{i} + 2C \, J_{\textrm{down}}\sum_{\langle j \rangle} s_{j} - \mu h \sum_{\langle j' \rangle} s_{j'} \,, \nn\\
  && \text { for any } s_{i}=+1 ~\text{and}~ s_{j,j'}=-1 \,.
\end{eqnarray}
The first sum runs over all pairs of nucleons with $s=+1$, and the second sum is over all pairs with $s=-1$. The factor $2$ is introduced to remind that there are two lattice sites with same spin in one pair, the dimension of coefficient $C$ is $E$, it is used to characterize the relative interaction strength compared to the external magnetic field. The third sum depicts the universal interaction with external magnetic field, we take $h$ to a negative value whose functionality only acts on the $-1$ states, reducing the energy of this system. This is consistent with the function of $U$ in the theoretical model introduced at previous section in Eq.\,(\ref{introU}).

One important feature of Metropolis algorithm shown in Eq.\,(\ref{MProb}) indicates that it allows MCS which increases the energy, albeit with a low probability if they increase the energy by a large amount. The variation of energy in every MCS is influenced by the couplings $J_{\textrm{up}}$ and $J_{\textrm{down}}$ which are expressed in terms of average Ising spin in this modified Ising model,
\begin{eqnarray}
\left\{\begin{aligned}
 & J_{\textrm{up}} = -\frac{1}{L^2} \sum_{\mu} s_\mu = -\bar s \,,\\
 & J_{\textrm{down}} = 1+\frac{1}{L^2} \sum_{\mu} s_\mu = 1+\bar s \,.
\end{aligned}\right.
\end{eqnarray}
Before simulation, we also need to specify the initial configuration, here it is all spins pointing down (i.e., $s_i=-1$ for all $i$) at $t=0$. This configuration corresponds to a bunch of ``free'' nucleons without SRC pairs. Then we add an external magnetic field and evolve this system until it reaches the equilibrium distribution. The simulation results would be shown in next section.

\subsection{Final results}

\begin{figure}[htbp]
\begin{minipage}{0.49\columnwidth}
\centerline{\includegraphics[width=1.00\columnwidth]{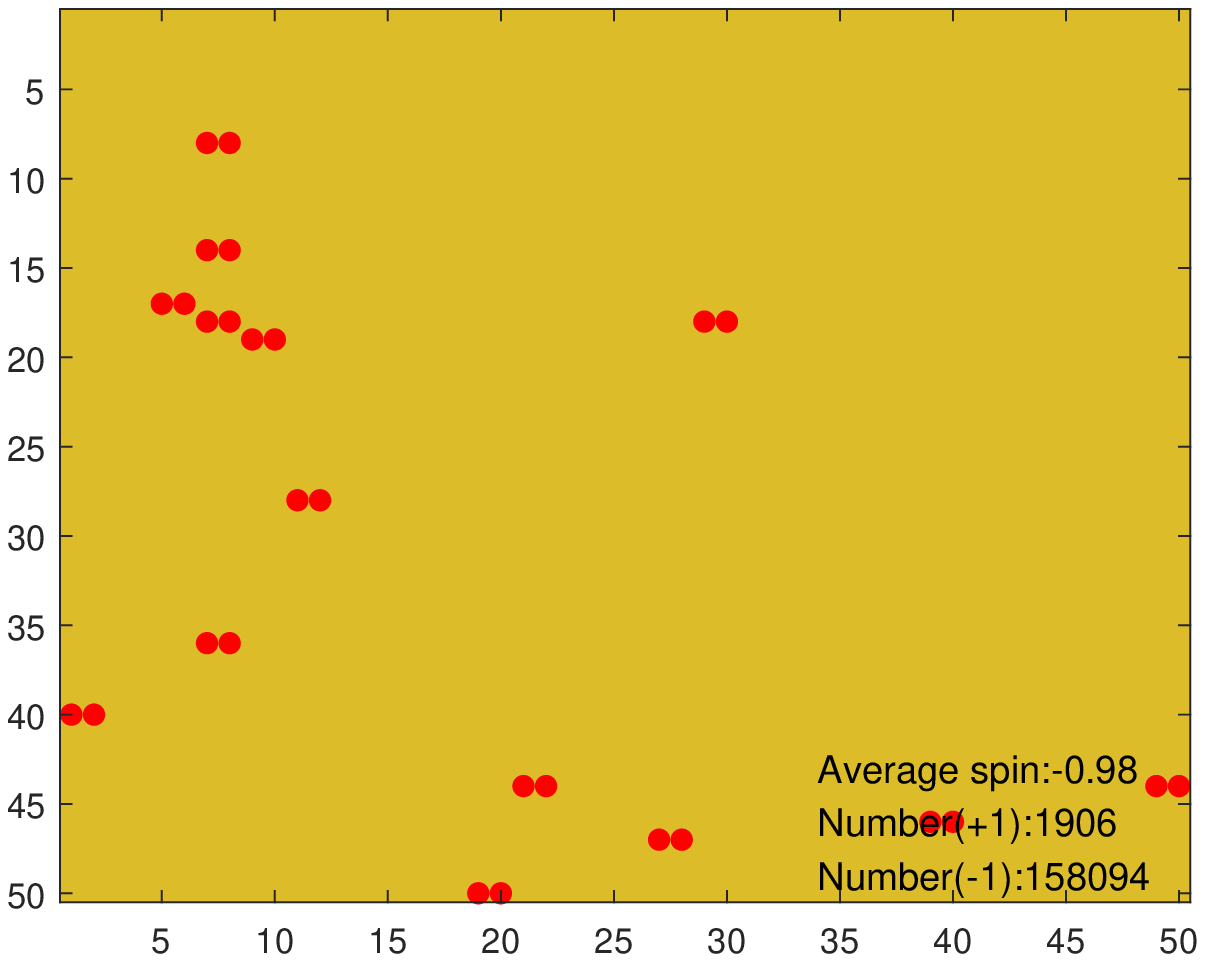}}
\vspace{0pt}
\caption*{(a)}
\centerline{\includegraphics[width=1.00\columnwidth]{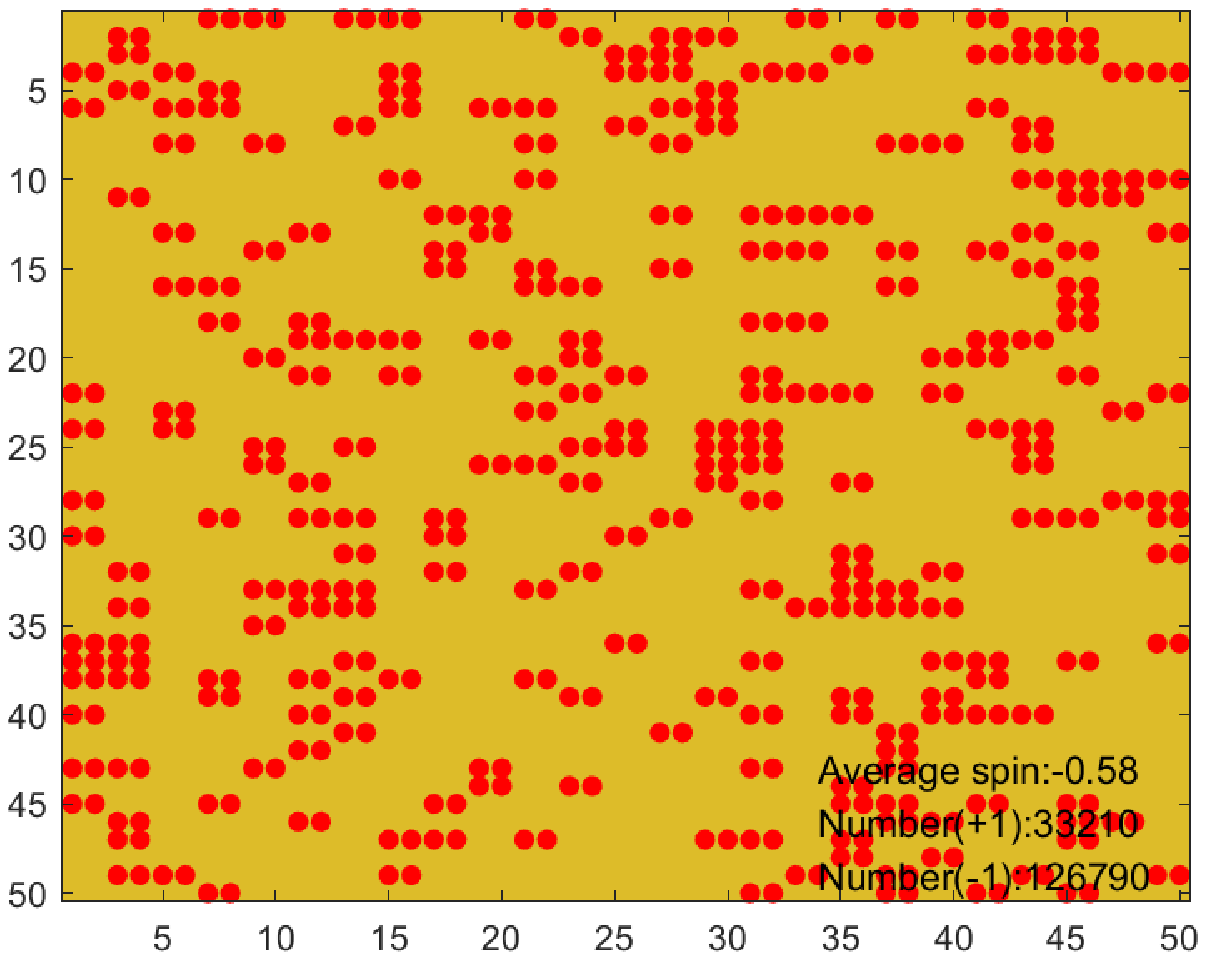}}
\vspace{0pt}
\caption*{(c)}
\end{minipage}
\begin{minipage}{0.49\columnwidth}
\centerline{\includegraphics[width=1.00\columnwidth]{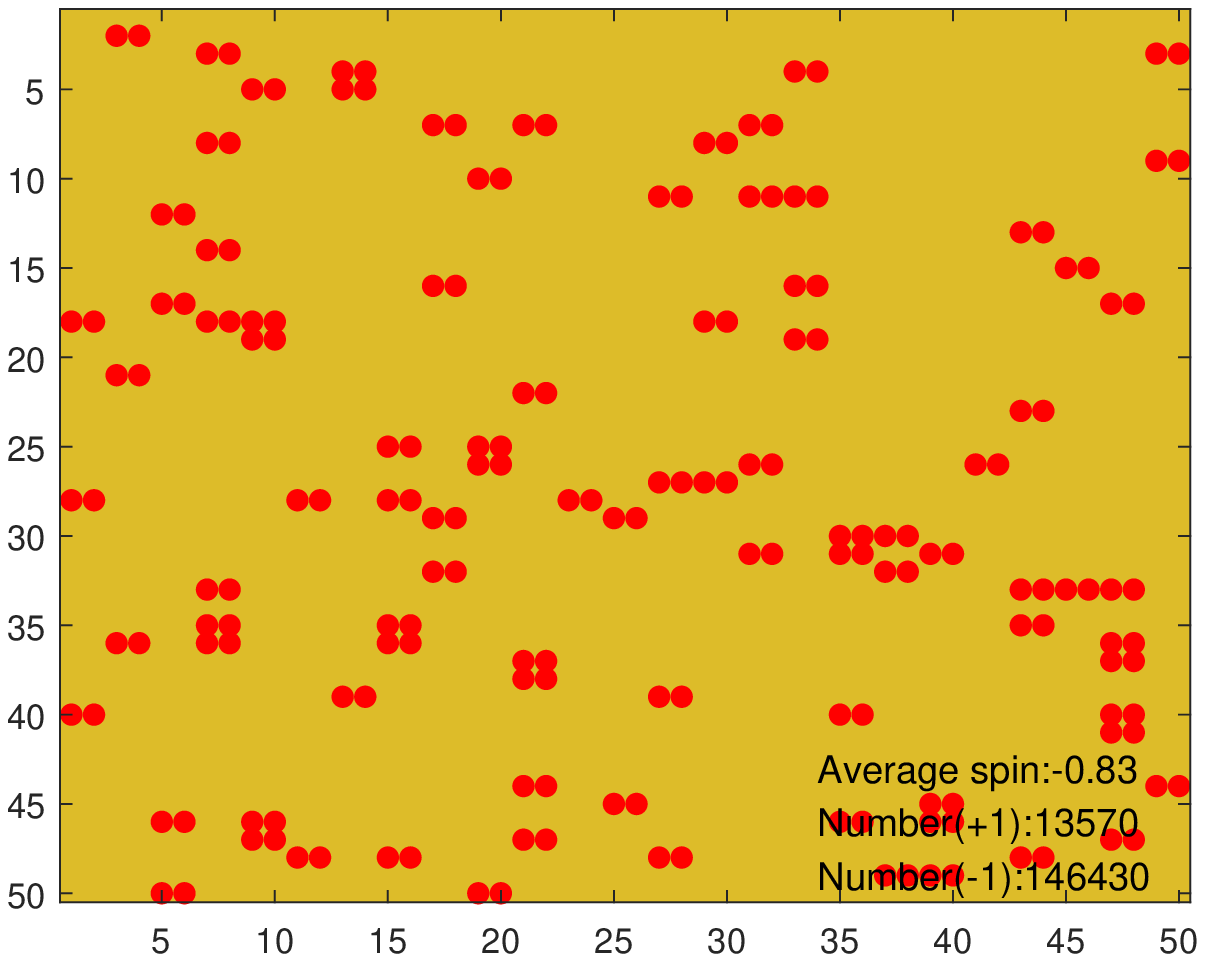}}
\vspace{0pt}
\caption*{(b)}
\centerline{\includegraphics[width=1.00\columnwidth]{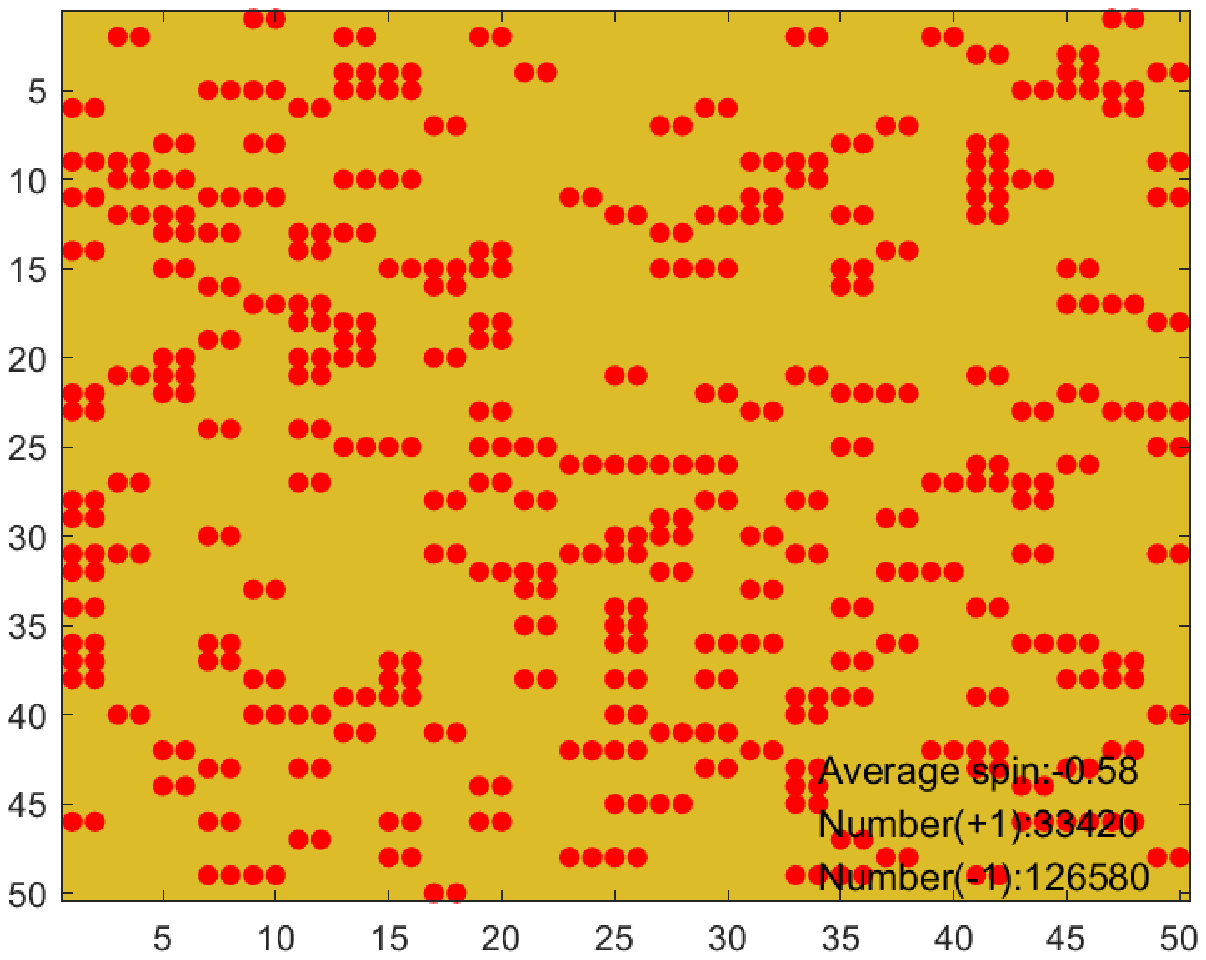}}
\vspace{0pt}
\caption*{(d)}
\end{minipage}
\caption{Simulation of bound nucleons in terms of two-dimensional lattice of size $400\times 400$ sites, here we only show a small part of the simulation micrograph ($50\times 50$) for better view. Every pair of red balls indicates a pair of SRC, other lattice sites tinted in yellow represent nucleons which are nearly free. The micrograph (a) is taken after completion of $1\times 10^3$ MCS in which the memory of initial configuration has not been lost. Micrograph (b) is taken after $1\times 10^4$ MCS where the variation of the system tends to be gentle. Micrograph (c) and (d) are taken after $5\times 10^5$ and $1\times 10^6$ MCS respectively, one can tell from these two diagrams that the system has reached equilibrium state. When $T=2.5$, the ratio of the two components maintains at $1/4$.}
\label{simulation}
\end{figure}

\begin{figure}[htbp]
\centering
\includegraphics[width=0.65\columnwidth]{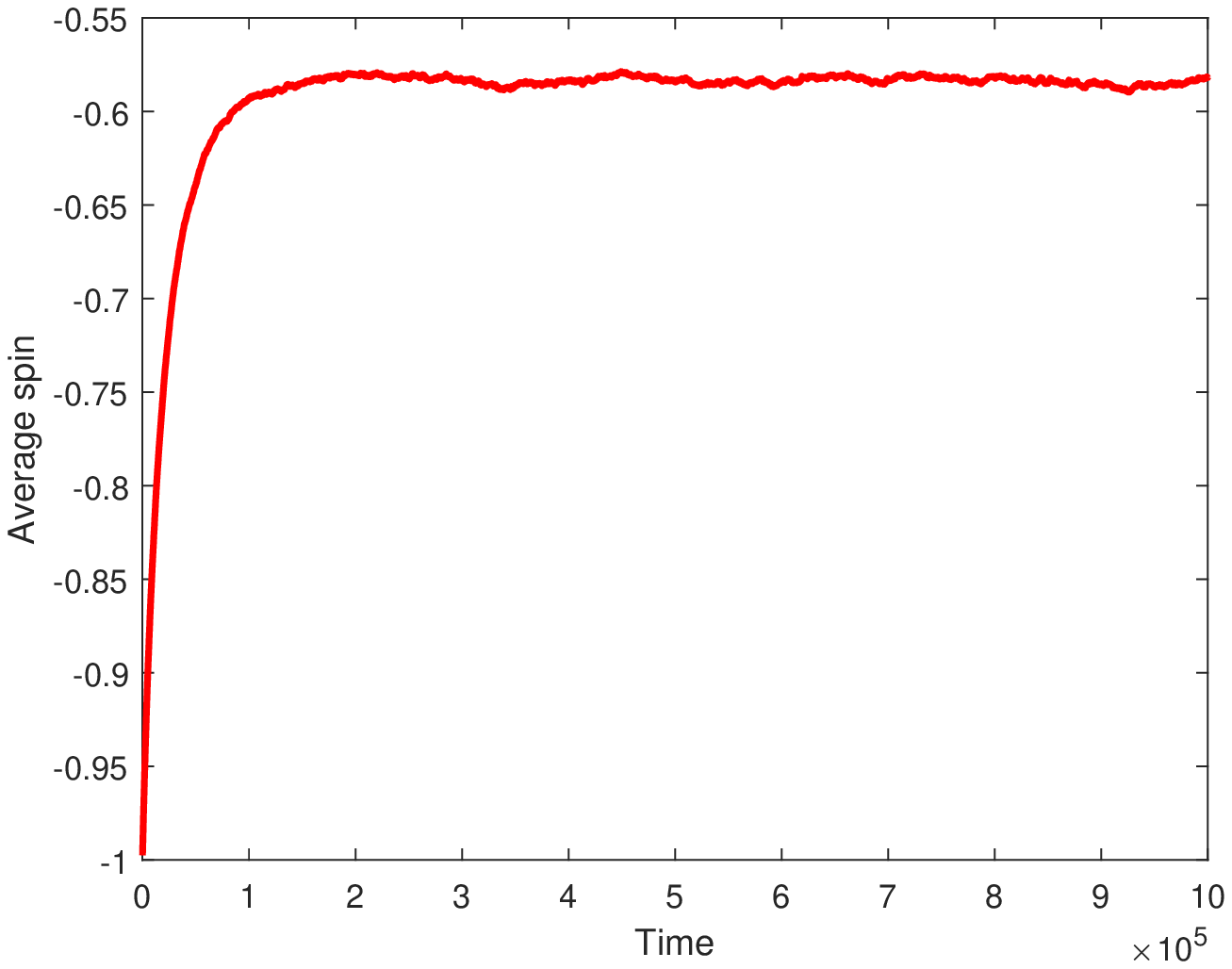}
\caption{Average spin vs time, it clearly shows the average spin of this system stabilizes at about $\bar s=-0.6$ after $1\times 10^5$ MCS, which is consistent with the micrographs in Fig.\,\ref{simulation}.}
\label{svstime}
\end{figure}

Here we present the final simulation results for the modified Ising model which are utilized to mimic the nucleons bound in a medium or heavy nuclei. Fig.\,\ref{simulation} presents the evolution of the system from initial state ($s_i \!=\! -1$ for all $i$) to equilibrium state. After completion of $1\times10^6$ MCS, the ratio of the two components (nucleon belongs to SRC or not) maintains at $1/4$ at $T=2.5$, which is consistent with experimental data. In this simulation, we take the coefficient $C=2$ and the external field $h=-4$. Fig.\,\ref{svstime} shows the stability of this simulation, after $1\times 10^5$ MCS, $\bar s$ remains at about $-0.6$.

\begin{figure}[htbp]
\centering
\includegraphics[width=0.65\columnwidth]{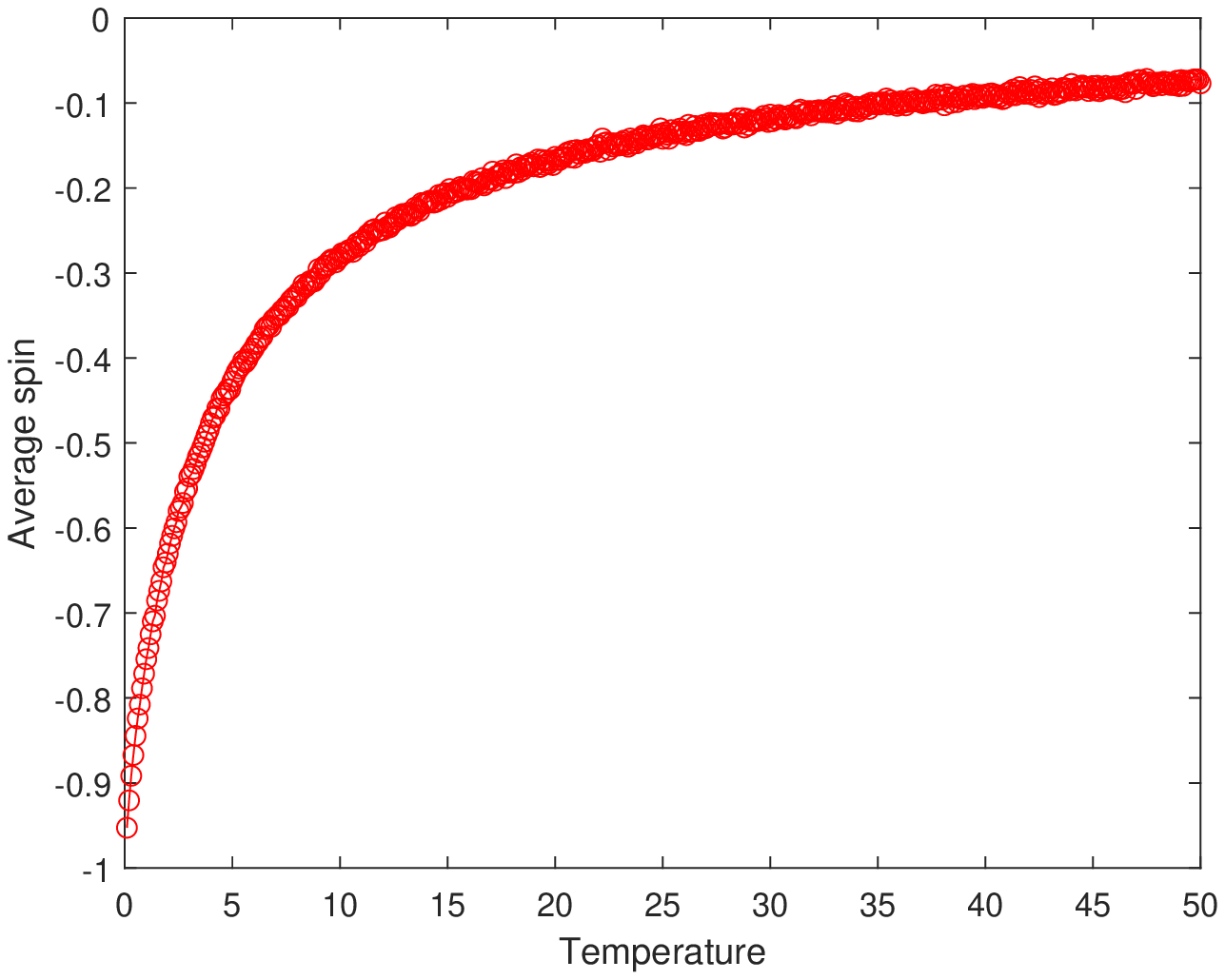}
\caption{Average spin vs temperature, the values are taken after $1\times 10^6$ MCS, here the $C$ and $h$ are fixed at $2$ and $-4$ respectively. $|\bar s|$ decreases with increasing temperature.}
\label{svsT}
\end{figure}

\begin{figure}[htbp]
\centering
\includegraphics[width=0.65\columnwidth]{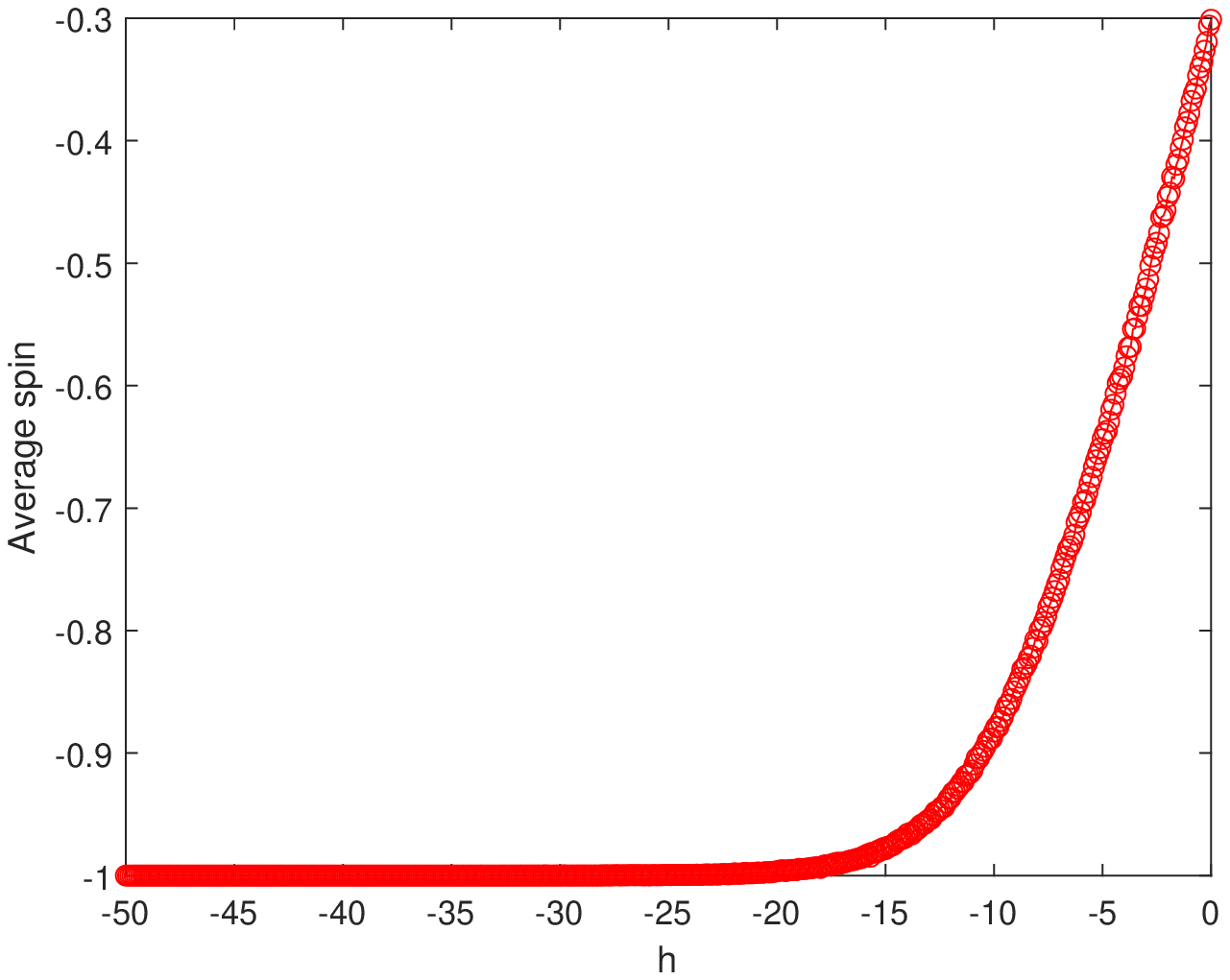}
\caption{Average spin vs external magnetic field, the values are taken after $1\times 10^6$ MCS, here we take $C=2$ and $h=-4$. $|\bar s|$ increases with increasing $|h|$.}
\label{svsh}
\end{figure}

Average spin $\bar s$ vs temperature $T$ curve is plotted in Fig.\,\ref{svsT}. As is evident from this figure, the system tends to more disordered (i.e., $\bar s \to 0$) as the temperature increases, whose corresponding situation in nucleus has been described in last section as $\Delta E^{-1} \to \infty$. Fig.\,\ref{svsh} presents the influence of external magnetic field $h$ on $\bar s$. When reaching the equilibrium state, the system is more orderly (i.e., $\bar s \to -1$) as $|h|$ increases.

The results shown in this section have qualitatively confirmed the validity of utilizing Ising model to describe the states of nucleons which are bound in nucleus. This encouraging result indicates the probability of investigating nucleons in terms of many well established techniques in thermodynamic statistics, which in some sense, assumes the fundamental mechanism of the real world is intrinsically probabilistic.

However, we want to note that the simulation in this manuscript is rather rough and the results are pretty sensitive to model constructions. The value of average spin will stabilize elsewhere instead of $-0.6$ as if the parameters in Eq.\,(\ref{M_Ising_H}) are changed (as shown in Fig.\,\ref{svstime} and Fig.\,\ref{svsT}). Therefore, to fulfill the power of Ising model, systematic studies on parameter selection are urgently called for. Besides, the estimation of $20\%$ nucleons which belong to SRC in medium or heavy nuclei needs to be explored in more detail, this Ising model based simulation should be able to describe a series of explicit nuclei and reproduce the linear relation between the magnitude of the EMC effect and SRC scale factor\,\cite{Weinstein:2010rt}.

\section{Conclusions}
\label{conclusions}
Based on the fact that the SRCs in nucleons and Ising spins in Ising model could share the same type of observables, a new SRC-bit map is proposed in this work. As a powerful tool, this map connects the state of each nucleon with state of each lattice site. We have considered a nuclear theory as a correspondence to an explicit Ising model and implemented a simulation of nucleon states in terms of Ising model, our preliminary results support the proposed map between nucleons and Ising spins.

Apparently, the investigation of nucleons by SRC-bit map is at the nascent stage, but with the advancements in computational sources and efficient algorithms, it's applications in research areas related to nuclear structure appear to be bright. More rigorous investigations on the related issue are urgently called for. Besides conceptual advances, the treatment of classical statistics and quantum particles in a common formalism could lead to unexpected cross-fertilization on both sides.

\section*{Acknowledgements}
We thank Prof. Wei Wang, Dr. Shuai Zhao and Dr. Jian-Ping Dai for valuable discussions. H.S.M and J.X. is supported in part by National Natural Science Foundation of China under Grant No. 12105247, the China Postdoctoral Science Foundation under Grant No. 2021M702957. Y.S.L is supported in part by National Natural Science Foundation of China under Grant No. 12002209.

\par\vskip40pt

\bibliographystyle{apsrev4-1}
\bibliography{bibliography}

\end{document}